# The jet–disc connection: evidence for a reinterpretation in radio loud and radio quiet active galactic nuclei

David Garofalo[1,2]

[1]*Department of Physics and Astronomy, Southern Polytechnic State University, Marietta, GA 30060, USA*

[2]*Department of Physics, Columbia University, New York, NY 10027, USA*

## ABSTRACT

To constrain models of the jet–disc connection, we explore Eddington ratios reported in Foschini (2011) and interpret them in relation to the values in Sikora et al. across the active galactic nuclei population from radio loud quasars, their flat spectrum radio quasar subclass, the recently discovered gamma-ray loud narrow-line type 1 Seyfert galaxies, Fanaroff–Riley type I (FR I) radio galaxies and radio quiet quasars of the Palomar Green survey. While appeal to disc truncation in radiatively inefficient flow appears to explain the observed inverse relation between radio loudness and Eddington ratio in radio loud and radio quiet quasars, FR I objects, scale invariance and recent data on powerful jets in narrow-line Seyfert 1 galaxies offer compelling arguments in favour of a reinterpretation of the jet–disc connection.

## 1 INTRODUCTION

Since the discovery of enormous energy production from quasars in the early 1960's, the search for explanation has led to a progressive refinement of the accreting black hole idea, culminating over the decades in the so-called spin paradigm (Shakura & Sunyaev 1973; Blandford & Znajek 1977; Blandford 1990; Wilson & Colbert 1995; Moderski, Sikora & Lasota 1998; Sikora, Stawarz & Lasota 2007). According to this picture, active galaxies are characterized by accretion on to a central supermassive black hole that in some cases involves energy release in the form of jets by some as yet unknown coupling between the disc and black hole (Blandford & Znajek 1977; Blandford & Payne 1982). When jet power empirically exceeds the power available in accretion, rotational energy from the black hole becomes the prime candidate for energy release in the form of electron/positron pairs (Penrose 1969; Williams 1995) and used to fuel a black hole jet that may in turn be collimated by a magnetodhydrodynamic disc jet (Blandford & Payne 1982; Hardee & Hughes 2003; Bogovalov & Tsinganos 2005). While confidence has grown in the idea of black hole accretion over the past four decades, our understanding of the jet–disc connection in the spin paradigm has been problematic. And certainly no promising explanation has emerged there for the quantitative nature of the radio loud/radio quiet division.

But the observations on the relation between discs and jets are not silent. The radio loud quasars (RLQ) on average have lower Eddington ratios compared to radio quiet quasars (RQQ) (Sikora et al. 2007) which seem to fit in the spin paradigm via high accretion rates in RLQ being associated with radiatively inefficient accretion flow in the inner regions where disc truncation allegedly sets in. However, this idea is problematic because it fails to be reflected in black hole X-ray binary behaviour, thereby breaking scale invariance. In addition, recent evidence has emerged showing that spiral galaxies can produce powerful jets (Komossa et al. 2006; Foschini 2011), which requires high black hole spin in the spin paradigm, providing us with a wonderful opportunity to constrain the jet–disc connection. In this light, we identify a statistically significant difference between Eddington ratios of Gamma-ray loud Narrow-Line Seyfert 1 galaxies (Γ-NLS1) and flat spectrum radio quasars (FSRQ), a feature that is at odds with simple spin paradigm expectations. In Section 2, we explore this in FSRQ and Γ-NLS1, highlighting the puzzling nature of this difference. But differences within the two populations also provide us with valuable clues which we explore in Section 3 addressing FSRQ and Γ-NLS1 separately. Here, we show that both populations display jet power inversely related to disc efficiency, but in a way that avoids the low Eddington ratio/high radio loudness region of the Eddington ratio/radio loudness plane for Γ-NLS1, and beyond the explanatory confines of disc truncation models. In Section 4, we explore specific classes of active galactic nuclei (AGN) in the context of both the spin paradigm and the recent gap paradigm, arguing for a reinterpretation of the jet–disc connection. In Section 5, we conclude.

## 2 EDDINGTON RATIOS IN FSRQ VERSUS Γ-NLS1

We explore the data reported in Foschini (2011) table 1 which involves FSRQ and Γ-NLS1 (Abdo et al. 2009; Ghisellini et al. 2010). The focus is on Eddington ratio disc luminosity and radio loudness parameter.

| Source | $L_{disc}/L_{Edd}$ | Radio loudness parameter |
|---|---|---|
| 3C 273 | 0.4 | 558 |
| 3C 279 | 0.025 | 75558 |
| 3C 454.3 | 0.2 | 30610 |
| B2 1520+31 | 0.015 | 18441 |
| B2 1846+32A | 0.13 | 2434 |
| B3 0650+453 | 0.1 | 5302 |
| B3 0917+449 | 0.2 | 9427 |
| B3 1633+382 | 0.1 | 20254 |
| PKS 0227-369 | 0.1 | 6955 |
| PKS 0347-211 | 0.1 | 13560 |
| PKS 0454-234 | 0.05 | 8452 |
| PKS 1454-354 | 0.15 | 6236 |
| PKS 1502+106 | 0.13 | 13397 |
| PKS 2023-07 | 0.05 | 9480 |
| PKS 2144+092 | 0.1 | 6490 |
| PKS 2201+171 | 0.04 | 9532 |
| PKS 2204-54 | 0.18 | 6926 |
| PKS 2227-08 | 0.11 | 61576 |
| PKS B0208-512 | 0.14 | 16954 |
| PKS B1127-145 | 0.25 | 3101 |
| PKS B1508-055 | 0.2 | 1752 |
| PKS B1510-089 | 0.04 | 3960 |
| PKS B1908-201 | 0.2 | 26215 |
| PMN J2345-1555 | 0.06 | 4485 |
| S3 2141+17 | 0.12 | 273 |
| S4 0133+47 | 0.1 | 13646 |
| S4 0954+55 | 0.02 | 4517 |
| S4 1030+61 | 0.04 | 4832 |
| S4 1849+67 | 0.05 | 17548 |
| SBS 0820+560 | 0.15 | 11236 |
| 1H 0323+342 | 0.9 | 40 |
| FBQS J1102+2239 | 0.4 | 13 |
| PKS 1502+036 | 0.8 | 1926 |
| PKS 2004-447 | 0.2 | 4198 |
| PMN J0948+0022 | 0.4 | 1153 |
| SBS 0846+513 | 0.4 | 1937 |
| SDSS J1246+0238 | 0.76 | 102 |

Table 1: FSRQ top group, Γ-NLS1 bottom group.

The objects in Table 1 are plotted in Fig. 1. The average Eddington ratio for the Γ-NLS1 is 0.56 while that for the FSRQ is 0.11 with a statistically significant difference of more than 4 standard errors. This result is puzzling. According to the spin paradigm, the two populations differ only in scale with FSRQ having larger black hole masses on average. Therefore, on average their Eddington ratios would be indistinguishable. To what extent can we stretch the spin paradigm to accommodate this data under the assumption that the two populations are in fact scale-invariant equivalent? From the expression for disc luminosity via

$L_d = \eta \dot{M} c^2$,

(with $\eta$ the disc efficiency and $\dot{M}$ the accretion rate) the observed Eddington ratios in Table 1 can be obtained by considering a range of possible values in both $\eta$ and $\dot{M}$, the only two variables we can tune. Our goal is to explore whether the range in these variables that we can appeal to is compatible with scale invariance. But the range in $\eta$ is quite limited which forces a limited range in accretion rate. Fig. 2 shows the accretion luminosity for fixed accretion rate for a Shakura & Sunyaev accretion disc. The disc luminosity depends on the inner boundary, which is chosen as the innermost stable circular orbit, a location that depends on black hole spin. We see that the value of luminosity varies by less than a factor of 10 over the entire spin range from high retrograde to high prograde. The relativistic calculation produces an even smaller range of efficiencies so we will ignore it here in order to produce upper limits on our theoretical models. In short, Fig. 2 shows us that over the entirety of its possible values, $\eta$ is rather limited in range. But the spin paradigm restricts us to an even smaller range in $\eta$ due tothe fact that it prescribes high prograde spins for both FSRQ and $\Gamma$-NLS1 objects, the reason being the powerful jet. Fixing the spin to high prograde values, therefore, forces us to appeal to differences in Eddington accretion rates between the two populations in order to explain their observed Eddington ratios. Apparently, we need larger Eddington accretion rates on average in $\Gamma$-NLS1 compared to FSRQ and this means an appeal to higher Eddington accretion rates by a factor of 5. On its face, this seems to be incompatible with models of galaxy formation, which propose that highest rates of cold accretion result from mergers, which are dynamically less relevant in spirals. Even if such objects are subject to recent mergers as has been proposed (Mathur 2000) such as in 1H 0323+342 (Anton, Browne & March 2008), we might expect accretion rates to be similar to those in FSRQ. But why an average increase by a factor of 5? Appeal to higher Eddington accretion rates in $\Gamma$-NLS1 compared to FSRQ does not, therefore, seem to be the most naturally appealing strategy.

But our failure to rescue the spin paradigm in the context of Shakura & Sunyaev accretion suggests that we appeal to different accretion scenarios. In fact, there is a body of theory for which powerful jets are compatible with outer thin discs associated with disc truncation in the inner regions where geometrically thick flows enhance the magnetic fields used to propel the jet (Meier 2001). While the range in accretion efficiency that we can appeal to is larger in truncated disc models, arguments suggest that it is not amplified more than the entire span of Fig. 2. In fact, interpreting the evidence in FSRQ in terms of disc truncation, suggests radial values of disc truncation at 9 to 10 gravitational radii (Kataoka et al. 2007; Sambruna et al. 2009). In other words, the assumption of disc truncation due to radiatively inefficient accretion in these sources lowers the radiative efficiency in such a way as to make it compatible with the radiative efficiency of a standard Shakura & Sunyaev disc but with maximum spin in a retrograde configuration. Hence, for truncated discs, the range in $\eta$ that we can appeal to for a given high prograde black hole spin is about a factor of 10. In short, assuming a fixed high spin in the two groups but in the context of disc truncation no longer constrains us to as narrow a range in accretion in order to explain the observed Eddington ratios. Therefore, we can rescue the spin paradigm by arguing that disc efficiency in $\Gamma$-NLS1 is larger on average than in FSRQ by a factor of 5 and this would also be compatible with an absence of lower Eddington accretion rates in FSRQ. This is because models of high accretion are associated with lower Eddington luminosity in radiatively inefficient flow. Therefore, lower Eddington accretion rates (on average) in Γ-NLS1 compared to FSRQ – compatible with galaxy formation models – are also compatible with greater Eddington ratio in the Γ-NLS1. And the larger radio loudness parameters in FSRQ would also be compatible with this picture. It may be worth emphasizing that both groups involve efficient fuelling mechanisms, i.e. we are well outside the range of parameters where advection dominated accretion due to sub-Eddington accretion plays a role. Therefore, all this is consistent with the fact that the general class of NLS1 objects ere efficiently fuelled (Deo, Crenshaw & Kraemer 2006).

In summary, truncated inner discs appear capable of explaining the difference in Eddington ratio between Γ-NLS1 and FSRQ, the difference being higher Eddington accretion rates and therefore lower disc efficiency in the latter. However, as we take a more global approach, we will argue that our appeal to disc truncation ultimately fails and that simplicity warrants a reinterpretation of the data that does away with the concept of disc truncation.

## 3 EDDINGTON RATIOS AND RADIO LOUDNESS WITHIN THE FSRQ AND Γ-NLS1 GROUPS

By the standards of the basic spin paradigm in the absence of disc truncation – everything else being equal – we expect to find that jet power increases along with accretion power as the black hole spin increases. However, this trend seems to be violated. Let us divide the FSRQ objects of Foschini (2011) into those with radio loudness parameter above 9000 and those below that threshold and see how the Eddington ratios behave in the two groups. The choice of this value allows for a balanced number of sources in both groups but is otherwise arbitrary. The group with larger $R$ has an average observed Eddington ratio of 0.10 with a standard error of 0.02 while the FSRQ group with $R$ values below 9000 have observed Eddington ratio average of 0.13 and a standard error of 0.02. Despite an absence of statistical significance in the two data sets, insofar as there is a trend, it involves an inverse relation between radio loudness and observed Eddington ratios as is the case for the Γ-NLS1 objects. As mentioned, we can rescue the simple spin paradigm by including disc truncation. Accordingly, the more radio loud objects have higher accretion rates, which lead to larger inner disc truncation radius and thus lower disc efficiency. The lower radiative efficiency, in turn, produces thicker inner disc geometry that generates a more effective jet. Because we have just developed tools that so far have rescued the spin paradigm, we attempt to use this strategy to explain differences within the Γ-NLS1. However, we will find that an additional element beyond disc truncation is needed to produce compatibility between data and spin paradigm.

Here, we compare objects in the class of Γ-NLS1 in light of their observed radio loudness and Eddington ratios (Fig. 3). Note that the most radio loud object in the group is PKS 2004−447 with an Eddington value of 0.2 and a radio loudness parameter of 4198. At the other extreme, the highest Eddington ratio (0.9) is associated with one of the lowest radio loudness parameters at 40 (1H0323). Adopting the spin paradigm notion that high black hole spin exists in all Γ-NLS1 (i.e. fixing the spins within a small range), explaining the value of the radio loudness parameter tempts us to conclude that PKS 2004−447 has an accretion rate that is two orders of magnitude larger than in 1H0323 (to make up for the radio loudness parameter being two orders of magnitude smaller in the latter). In fact, in advection-dominated models, the square of the magnetic field value threading the black hole, and thus jet power, is linearly related to the accretion rate (Nemmen et al. 2007). Yet the Eddington ratio in 1H0323 is about a factor of 5 larger compared to PKS 2004−447, which makes this strategy problematic. In fact, observed Eddington ratios that vary by only a factor of 5 are beyond the obtainable values for Eddington accretion rates that differ by a factor of 100. In other words, the spin paradigm requires incompatible behaviour in the accretion rates in order to satisfy both the observed Eddington ratios and radio loudness parameters. Again, as we strategized above, let us appeal to differences in disc efficiency, the only other free parameter. In order to produce a larger Eddington ratio for 1H0323, implies a higher black hole spin in 1H0323 compared to PKS 2004−447. But we now find it difficult to explain the radio loudness difference between the two.

For groups with non-negligible jet power, the space of possible jet powers associated with possible accretion rates, and possible disc efficiency, becomes fairly tightly constrained. In particular, for multiple sources whose accretion efficiencies and Eddington accretion rates live within a narrow range due to a given observed Eddington ratio, their jet powers must also live within a relatively narrow range. If we fix the accretion rate, differences in jet power come from black hole spin, which is proportional to the spin value squared in the Blandford–Znajek mechanism. Numerical simulations offer differences in spin dependence from Blandford–Znajek but such differences in power between objects with non-negligible jets vary by a factor of a few. For example, take two objects whose spin values are 0.5 and 1; the jet power differs by a factor of 6 (Tchekhovskoy, McKinney & Narayan 2012). The point is that in the context of the spin paradigm for objects with powerful jets, there are no large variations in jet power for small variations in parameters that determine the Eddington ratio. And because we argued that disc efficiency is still constrained to be within one order of magnitude in radiatively inefficient flow, this argument extends to disc truncation models. Objects that drastically violate this are difficult to reconcile with the spin paradigm. In order to explain the observations for the Γ-NLS1, we seem to need a mechanism that can produce drastic differences in jet powers for small differences in the parameters that generate the Eddington ratio but in a way that allows Γ-NLS1 to occupy regions of the Eddington ratio versus radio loudness plane (upper left of Fig. 1) that is not accessible to FSRQ (lower right of Fig. 1). And this mechanism should be scale free. The bottom line in our analysis is that observed values of Eddington ratios and radio loudness parameters produce constraints on the jet–disc connection. In the next section, we identify a scale-invariant mechanism that addresses these issues.

## 4 A REINTERPRETATION OF THE JET–DISC CONNECTION

According to the gap paradigm (Garofalo, Evans & Sambruna 2010), FSRQ have lower disc efficiency compared to Γ-NLS1 due to thin-disc accretion in a retrograde accretion configuration, with large gap regions resulting from a missing reservoir of potential accreting material in the inner black hole region compared to their prograde counterparts. Because Γ-NLS1 appear to live in spiral galaxies, and such galaxies struggle to produce retrograde accretion due to their smaller and thus less stable retrograde black hole configurations, they are modelled as prograde accreting systems in the gap paradigm. However, if the accretion is radiatively efficient, scale invariance arguments require that jets form only in a narrow range of intermediate prograde spin values (Garofalo 2013). But intermediate prograde accreting systems have disc efficiencies whose values are sandwiched between the retrograde objects (at lowest $\eta$) and the high prograde objects (at highest $\eta$) as seen in Fig. 2. Therefore, FSRQ and Γ-NLS1 fit into a theoretical framework that does not require additional assumptions beyond standard radiatively efficient Shakura & Sunyaev accretion.

The reinterpretation of the jet–disc connection that is being suggested involves a tug of war between jets and discs. When gap regions between black holes and accretion discs are large (retrograde configurations), jets are most effective, uninhibited by disc-quenching, stemming from the relative weakness of disc winds. On the other hand, when gap regions are small (prograde configurations), jets *would* be weaker than in the retrograde scenario but would nonetheless still be present were it not for jet-quenching. In fact, the ability to quench jets by accretion (Neilsen & Lee 2009) depends on the strength of the disc wind, which increases when gap regions are smallest due to the larger disc efficiency (Garofalo et al. 2010). Therefore, when black hole spin is in the high prograde regime, disc-quenching of jets dominates the dynamics and such objects are RQQ-like. In short, the time evolution of a radiatively efficient disc that begins in a high spin but retrograde configuration involves the presence of powerful jets accompanied by weaker disc winds, but that gives way to a waning of the jet due to increased disc efficiency and wind power, and thus jet-quenching, as the system transitions towards the high prograde regime. This simple picture is not only compatible with the observations presented here in FSRQ and Γ-NLS1, but also with Fanaroff–Riley type I (FR I) radio galaxies and PG quasars as discussed further below.

Let us re-evaluate the observations in light of these ideas. Assuming the Eddington accretion rates to be about the same in both 1H0323 and PKS 2004−447, the observed Eddington ratios imply that 1H0323 has larger disc efficiency and thus the higher black hole spin. But the gap paradigm prescribes that as the spin increases in the prograde direction, jet quenching begins to dominate the dynamics, so the radio loudness parameter of 40 (i.e. two orders of magnitude less than in PKS 2004−447) seems no longer problematic. In other words, the system has crossed the prograde spin value for which thermal discs sustain powerful jets. The lower disc efficiency in PKS 2004−447, on the other hand, allows the disc to experience a relatively larger gap region (because the spin value is lower), which allows the jet to remain unquenched, i.e. the threshold for jet-quenching has not been crossed. Therefore, the gap paradigm predicts that if the Eddington accretion rates are sufficiently similar in the two objects, the difference must be in disc efficiency, which depends on black hole spin, larger for 1H0323. Similar arguments hold for the other objects. PKS 1502, for example, would fit in the gap paradigm as a system whose accretion rate is making up for the fact that it has low disc efficiency due to lower prograde spin. That would be why the radio loudness parameter is high. The large variation in radio loudness between 1H0323 and PKS 1502 despite similar observed Eddington ratios, coupled with the limited range of accretion rate on to the black hole allowed by compatibility with the observed Eddington ratio, suggests that such objects are giving us information on the threshold value of prograde black hole spin associated with jet quenching. In other words, the gap paradigm prescribes that at some intermediate value of prograde spin, thermal discs will become effective jet quenchers, so there should be a location where disc efficiencies vary by a small amount yet differences in jet power are more pronounced. In short, we have explained not only the statistically significant difference between Γ-NLS1 and FSRQ and thus their distribution on the Eddington ratio/radio loudness plane, but both the inverse relation between jet power and Eddington ratios as well as the large variations in jet power observed in Γ-NLS1 for small variation of disc parameters, all within the context of Shakura & Sunyaev-like accretion. It is important to point out that accretion rates above Eddington and the radiative inefficiency of the inner discs that result from this in models that extend beyond Shakura & Sunyaev, can still be reconciled with the observations as long as the disc spectrum and wind production from discs is still dependent on the disc inner edge, which in turn need not be coincidental with the innermost stable circular orbit, only monotonically related to it. The scale-invariant implication of this involves the necessity of associating soft X-ray binary states with slim disc type models (Abramowicz et al. 1988). The regime where high enough accretion rate washes away any dependence of disc parameters on the inner edge determines where radiatively inefficient models of super-Eddington accretion become problematic (Kawaguchi 2003). We do not explore this further here.

Let us move beyond thermal discs. In radiatively inefficient or advective dominated discs the jet-quenching ability fails, and the jet power adopts a flatter spin dependence resulting from decreasing Blandford–Payne jet power with increase in prograde spin and increasing Blandford–Znajek jet power with increase in spin (Garofalo et al. 2010, fig. 3). This flatter spin dependence of jet power is also compatible with recent results from black hole X-ray binary jets in bright hard states where observations of jets associated with different black hole spins do not have very different powers (Fender, Gallo & Russell 2010). If the brighter hard state jets in X-ray binaries are the small-scale counterpart to FR I radio galaxies, this observational feature is radically at odds with the spin paradigm in all its forms, including analytic, semi-analytic and numerical simulations.

The trends observed above in the RLQ population are not new. Sikora et al. (2007) have shown that the radio loudness *R* increases with decreasing Eddington ratio for both RLQ and RQQ (see their fig. 3). But this trend does not occur for the FR I radio galaxies. If anything, the radio loudness parameter for FR I objects has a slight positive dependence on Eddington ratio (Sikora et al. 2007). This suggests an intrinsic difference at least between FR I objects and FSRQ, Γ-NLS1 and RQQ. This difference is interpretable in the gap paradigm as the result of absence of disc quenching due to radiatively inefficient flow in FR I objects, which allows the jet power to increase with prograde black hole spin but with a flatter dependence compared to the retrograde regime (Garofalo et al. 2010, fig. 3). But FR I, Γ-NLS1 and RQQ would all be prograde accreting objects. Only the RLQ would be retrograde objects. How do RLQ obey scale invariance?

Very recent work has suggested a new interpretation of the jet–disc connection, appealing to scale invariance in using very high state, transitory burst jets in microquasars, as the small-scale counterparts to RLQ and broad line radio galaxies by using trun-cated/refilling accretion discs (Lohfink et al. 2013; see also Punsly & Rodriguez 2013). However, as discussed in Garofalo (2013), any straightforward attempt at using microquasars to model RLQ implies scaling-up the time dependence of stellar-mass black holes and thus fails to be compatible with the redshift dependence. Why would truncated/refilling discs display redshift dependence in the AGN regime that is not observed in the average time dependence of X-ray binary state transitions? This issue does not arise in the gap paradigm since FSRQ and broad line radio galaxies are modelled as retrograde accreting black holes, a feature that is relevant only in systems whose ratio of black hole mass to accretion mass is high. Without additional assumptions, the gap paradigm naturally incorporates redshift dependence in RLQ and broad line radio galaxies (higher on average) compared to FR I radio galaxies, Γ-NLS1, RQQ and Seyfert galaxies, due to the fact that continued accretion will inevitably lead to prograde spins.

## 5 CONCLUSION

In light of observed Eddington ratios in RLQ, RQQ, Γ-NLS1 and FSRQ, and proxies for jet efficiency, we have argued that meaningful trends emerge among the different groups that constrain the jet–disc connection. In particular, we have shown that large differences in jet powers for small differences in accretion parameters are difficult to reconcile with the spin paradigm. A narrow range in spin values implies a narrow range in disc efficiency, which in turn implies a narrow range in Eddington accretion rates in order to generate the observed Eddington ratio. But we have found objects among the Γ-NLS1 class in which similar observed Eddington ratios are associated with drastically different radio loudness parameters. If the latter can be used as a proxy for jet power, the data is difficult to interpret within the spin paradigm, even in the context of disc truncation. What appears to be needed is a mechanism that can produce a wide range of non-negligible jet powers for narrow range of accretion rate and disc efficiency, but which does so for the Γ-NLS1 at high Eddington ratio, thereby explicitly distinguishing between FSRQ and Γ-NLS1 and in a scale-invariant way. We have shown how the simple ideas in the gap paradigm provide an attractive scale-free framework for interpreting this data, thereby suggesting that disc truncation constitutes unnecessary patchwork. The model makes specific predictions: are Γ-NLS1 characterized by intermediate prograde black hole spins?

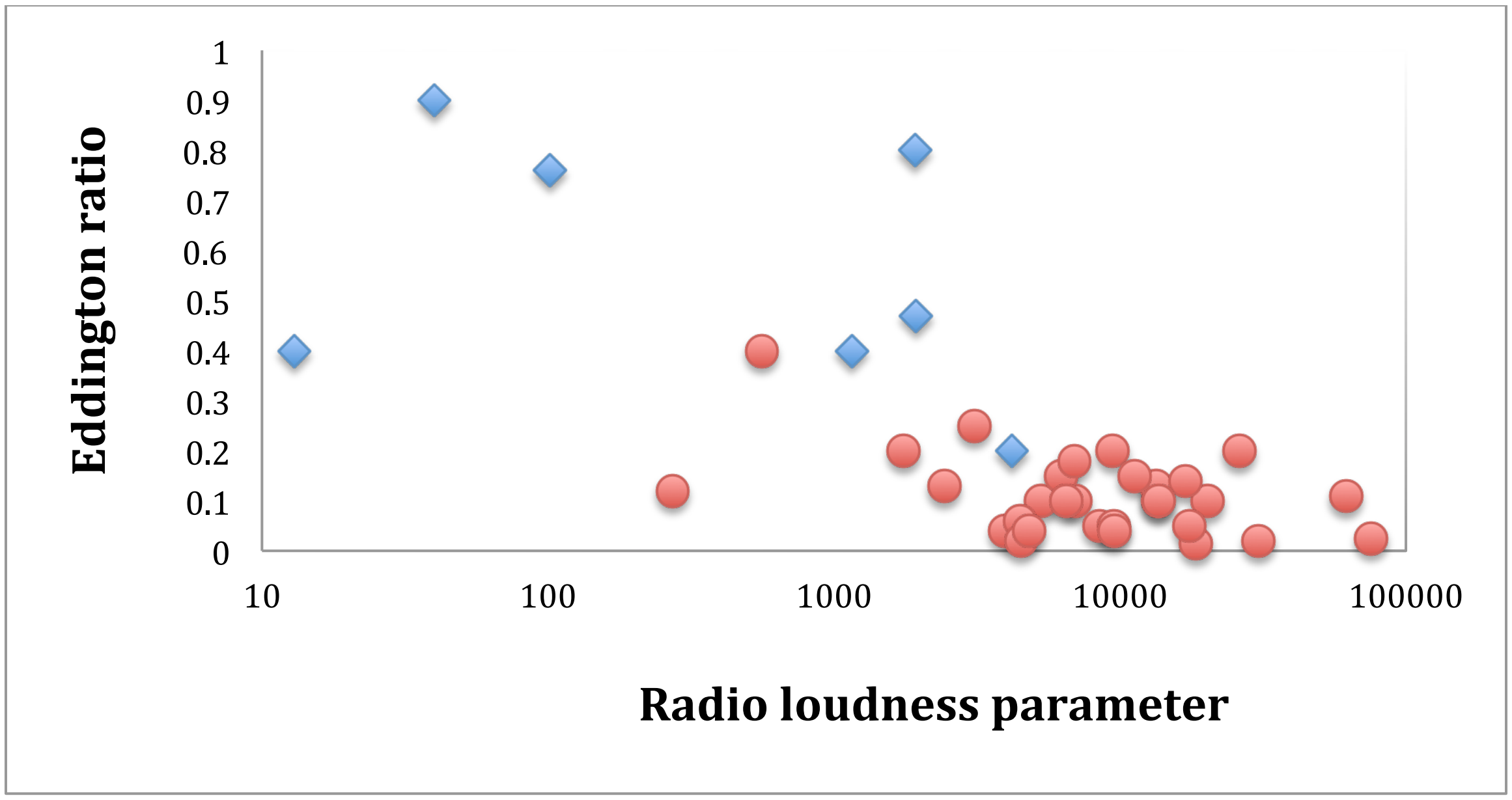

Figure 1: Eddington ratios vs radio loudness parameter for FSRQ (red) and Γ-NLS1 (blue). Data is from Table 1.

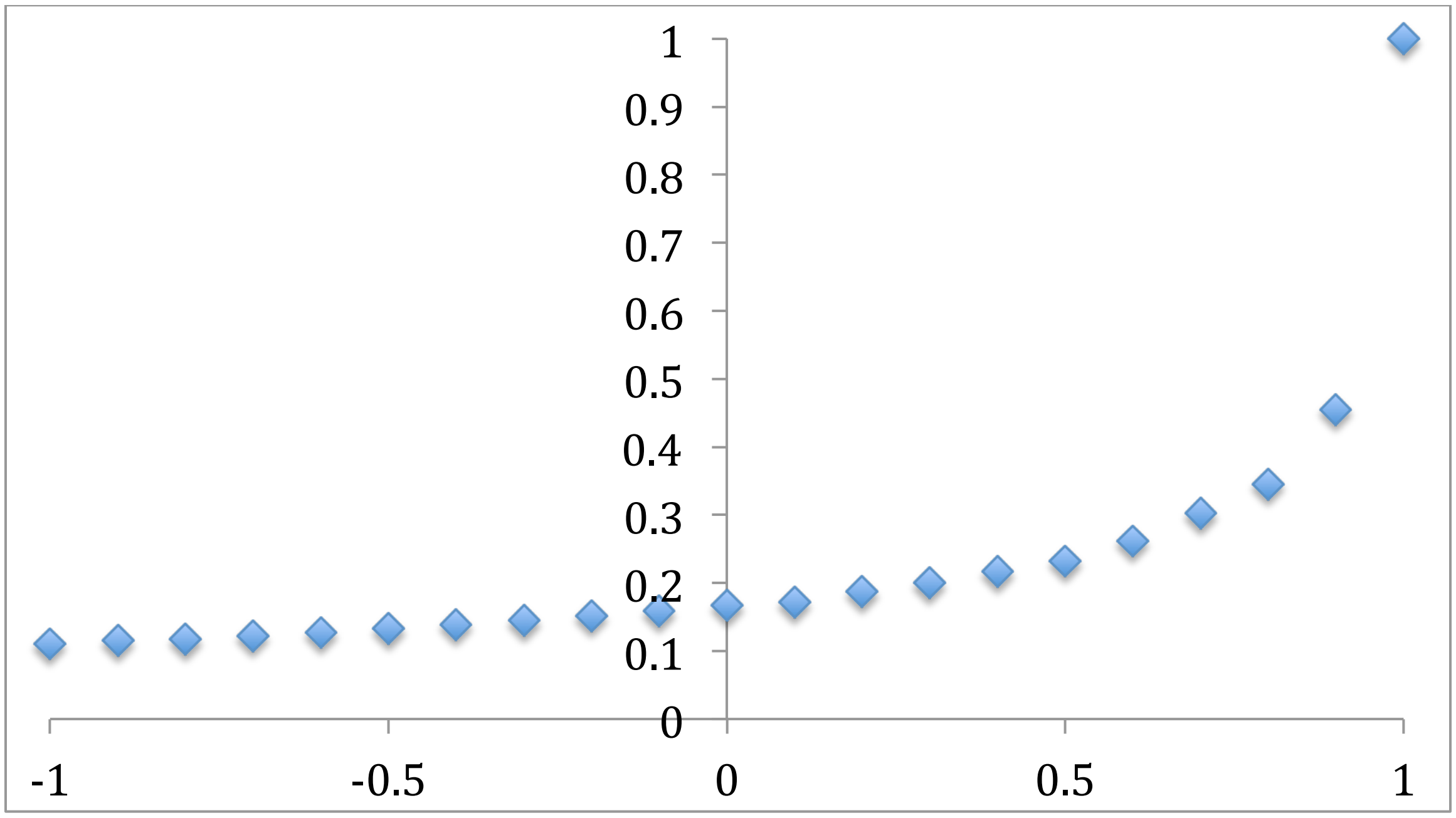

Figure 2: Disk luminosity for fixed accretion rate normalized to maximal prograde spin case vs spin for a Shakura & Sunyaev accretion disk. Retrograde disks are represented as negative x values while prograde disks live in the positive x range. The vertical axis can therefore be thought of as a normalized disk efficiency η.

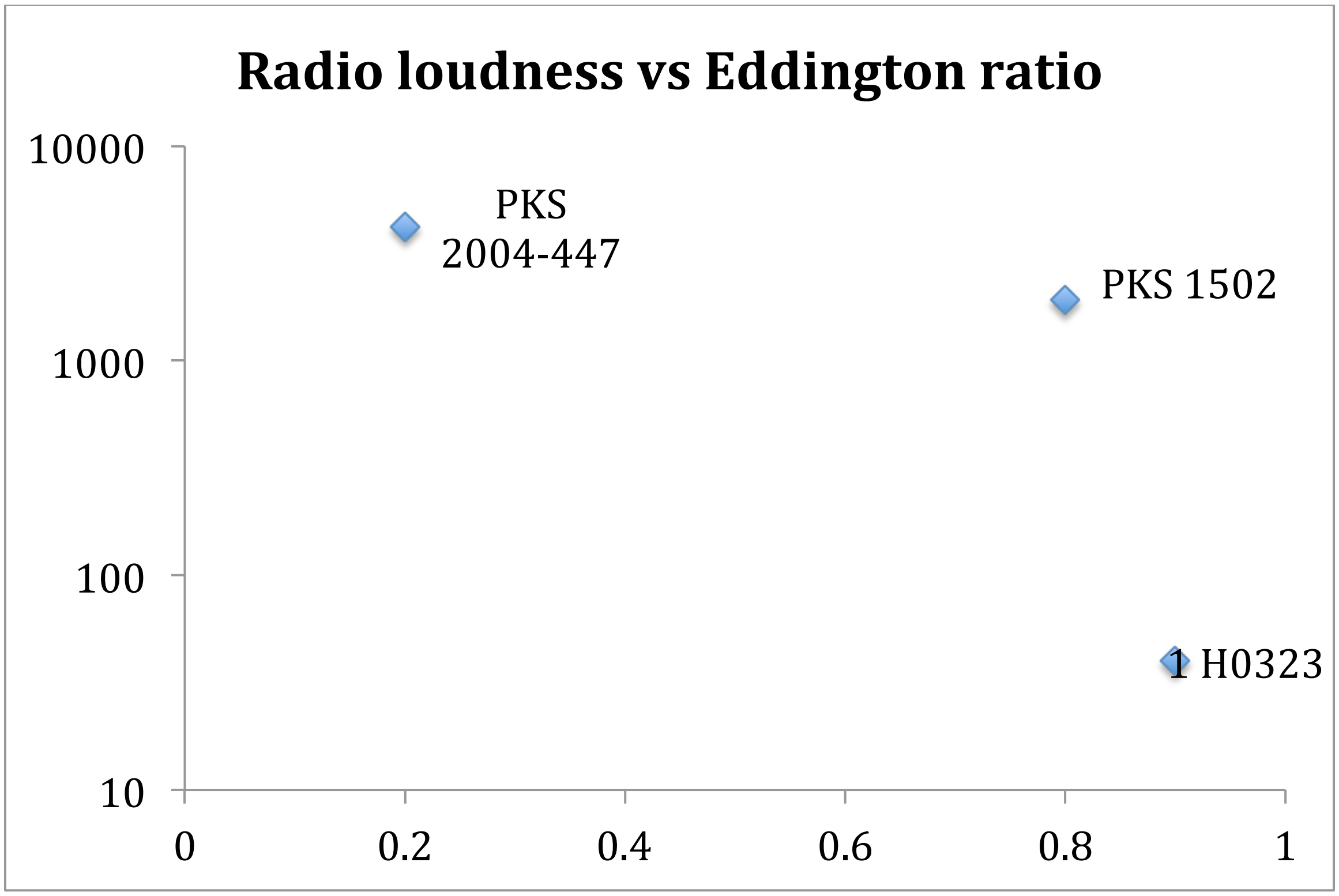

Figure 3 Three Γ-NLS1 objects viewed in the radio loudness–Eddington ratio plane. In the gap paradigm, higher disk efficiency for 1 H0323 due to higher prograde black hole spin is compatible with a lower radio loudness parameter compared to the other two sources. PKS 2004-447 would have the lowest prograde black hole spin among the three objects, which allows its jet to remain unquenched.

**ACKNOWLEDGEMENTS**

I thank Manel Errando and Reshmi Mukherjee for stimulating conversation and hospitality in the Veritas group. I also thank the referee of the first version of this paper who graciously went out of his/her way to emphasize the utility of specific figures, and the referee of the second version for strengthening my exploration of NLS1 galaxies.